\documentclass{article}
\usepackage{amsmath,latexsym,epsf}
\numberwithin{equation}{section}

\newcommand{\beq}{\begin{equation}}
\newcommand{\eequ}{\end{equation}}
\allowdisplaybreaks

\begin{document}

\begin{center}
\title{\bf Quantitative Predictions for ${\boldsymbol{B}}$ semileptonic decays into ${\boldsymbol{D}}$, ${\boldsymbol{D^\ast}}$ and the orbitally excited ${\boldsymbol{D^{\ast\ast}}}$  in quark models \`a la Bakamjian-Thomas}
\author{V. Mor\'enas$^a$ \\[5mm] 
 A. Le Yaouanc, L. Oliver, O. 
P\`ene and J.-C. Raynal$^b$}\par
\maketitle
{$^a$ Laboratoire de Physique Corpusculaire\\
Universit\'e Blaise Pascal - CNRS/IN2P3
F-63177 Aubi\`ere Cedex, France\\
$^b$ Laboratoire de Physique Th\'eorique et Hautes
Energies\footnote{Laboratoire
associ\'e au
Centre National de la Recherche Scientifique - URA D00063
\\  e-mail: morenas@clrcls.in2p3.fr, leyaouan@qcd.th.u-psud.fr, 
oliver@qcd.th.u-psud.fr}}\\ {Universit\'e
de Paris XI, B\^atiment 211, 91405 Orsay Cedex, France}
\end{center}

\begin{abstract}

Once chosen the dynamics in one frame, the rest frame in this paper, the
Bakamjian and Thomas method allows to define relativistic quark models
in any frame. These models have been shown to provide, in the infinite
quark mass limit, fully covariant current form factors as matrix
elements of the quark current operator. In this paper we use the rest
frame dynamics fitted from the meson spectrum by various authors,
already shown to provide a reasonable value for $\rho^2$. From the
general formulae for the scaling invariant form factors $\xi^{(n)}(w)$,
$\tau_{1/2}^{(n)}(w)$ and $\tau_{3/2}^{(n)}(w)$, we predict quantitavely the $B$
semileptonic branching ratios to the ground state and orbitally excited charmed mesons $D, D^\ast$ and $D^{\ast\ast}$. We check Bjorken's sum rule and discuss the respective contributions to it.
 We find $\xi(w)\simeq (2/(1+w)) ^2$, resulting  from the fact that the ground state wave function is Coulomb-like. We also find
 $\tau_{3/2}\simeq 0.5 (2/(1+w))^3$ and $\tau_{1/2}(w)\ll \tau_{3/2}(w)$.  Very small branching ratios into $j=1/2$ orbitally excited
$D$'s results.  The overall agreement with experiment is rather
good within the present accuracy which is poor for the orbitally excited charmed mesons.  We predict a ratio $Br(B\to D^\ast_2 l \nu)/Br(B\to D_1 l
\nu)=1.55\pm 0.15$ as a mere consequence of the heavy quark symmetry.  If some faint experimental indications that $Br(B\to D_1 l
\nu)\simeq Br(B\to D^\ast_2 l \nu)$ were confirmed, it would indicate a
sizeable $O(1/m_c)$ correction.

\end{abstract}
\begin{flushright} LPTHE Orsay-97/19 \\PCCF RI 9707 \\ hep-ph/9706265
\end{flushright}
\newpage

\section{Introduction}

The experimental progress in beauty physics these last years is really
astounding.  Subdominant decay channels such as the 
decay into non charmed final state are now currently seen and measured
with an increasing accuracy. In this paper, we are focusing our attention
on another class of non dominant decays, namely $B\to D^{\ast\ast}$.
Their suppression relative to the dominant $B\to D^{(\ast)}$ is of
course not due to CKM angles but to the dynamics of the decay, and
mainly to the orthogonality of the orbitally excited spatial wave
function of the $D^{\ast\ast}$'s to $B$'s the ground state one.

Among the four non-strange expected $D^{\ast\ast}$'s, two have been seen,
the $D^{*}_2(2460)^\pm$ with $I(J^P)=\frac 1 2 (2^+)$, and the
$D_{1}(2420)^\pm$ with $I(J^P)=\frac 1 2 (1^+)$, which have rather small
widths ($\simeq 20$ MeV). This small width is usually understood as a
consequence of being $j=3/2$ states, $j$ being the total angular
momentum of the light quanta (light quarks and gluons) of the
system\footnote{If the $D^{*}_2$ has to be a pure $j=3/2$, the two
expected $J=1$ states with $j=1/2, 3/2$ should mix due to $1/m_c$
corrections to the Heavy Quark Effective Theory (HQET). This mixing is
estimated to be small \cite{falk} and the observed state to be
dominantly $j=3/2$.}. The small widths of these states is attributed to
the centrifugal barrier effect on their decay. Indeed, the dominant
decay channels of the $D^{\ast\ast}$'s are $D^{(\ast)} \pi$ and
$D^{\ast\ast}(j=3/2) \to D^{(\ast)}(j=1/2) \pi$ is only allowed
through a $L=2$ partial wave by parity and $j$ conservation. Of course
$j$ conservation is only valid at leading order in $1/m_c$.  It
appears that the main expectations of HQET seem, up to now, to be rather
well fulfilled, even though the charm is not such a heavy quark and
important corrections would not be a surprise.  According to what seems
to be the standard use, we will denote the $j=1/2$ states:
$D^\ast_1(\sim 2420)$ with $I(J^P)=\frac 1 2 (1^+)$ and $D^\ast_0(\sim
2360)$ with $I(J^P)=\frac 1 2 (0^+)$.

There are altogether three main reasons which lead us to study these
$D^{\ast\ast}$'s with some care.  First, narrow resonances are in general
preferred for phenomenology for many practical reasons: they are
better isolated from the background and from other resonances, the very
convenient and frequent ``narrow resonance approximation'' can be
rightfully used, etc. Second, the apparent success of HQET at dominant
order in $1/m_c$ leads us to hope a reasonable success when describing
the $B\to D^{\ast\ast} l \nu$ decay in the infinite mass limit without
having to face the large number of new operators and unknown parameters
at next order in $1/m_c$. Last, but not least, the excited hadrons
represent a domain in which, to our knowledge, quark models remain the
only tool to make predictions, beyond general consequences from Heavy
Quark Symmetry (HQS). Lattice QCD as well as QCD sum rules meet huge
difficulties to deal with hadrons beyond the ground states.

Thus, $B\to D^{\ast\ast}$ is the privileged domain to test a class of
relativistic quark models \`a la Bakamjian and Thomas (BT) \cite{BT},
which we have recently proven \cite{raynal} to provide a covariant
description of semileptonic decays in the heavy mass limit.  It was
shown \cite{raynal} that this class of models satisfies fully HQS and
that the $\rho^2$ Isgur-Wise slope parameter was bounded in this class
of models: $\rho^2>0.75$. It also satisfies, \cite{duality},
\cite{dstar}, duality properties amounting in particular to the Bjorken
sum rule \cite{bjorken},
\cite{IW}. In  \cite{sr2} it is shown that they
satisfy the new sum rules involving annihilation constants that we have
proposed in
\cite{sr1}. 

The method of BT proceeds as follows. Given the wave function in, say,
the rest frame\footnote{Any starting frame can be chosen, for example
the infinite momentum frame although some care is then needed.}, the
hadron wave functions are defined in any frame through a unitary
transformation, in such a way that Poincar\'e algebra is satisfied. The
mass operator, i.e. the Hamiltonian operator in the rest frame, is not
specified in the BT approach. The mass operator is simply assumed to be
rotationally invariant, to depend only upon the internal variables (to
commute with total momentum and center of mass position) and, of course,
to conserve parity \cite{raynal}. In other words, the spectrum of heavy
flavored mesons is left free, and all the above mentioned properties are
valid for any spectrum.

These welcome properties make the BT method appealing enough to try
real phenomenology. The hadron spectrum has been studied many times, and
several good fits to the masses of heavy flavor mesons exist in the
literature. The question we ask is, {\it given any mass operator which
fits well the mass spectrum, what does it predict for the slope
$-\rho^2$ of the Isgur Wise function at $v.v'=1$, and for the $B\to
D^{\ast\ast} l \nu$ decay }? The results about $\rho^2$ have been
published and discussed in a separate publication \cite{rho2}. It has
been stressed that the detailed shape of the potential (more or less
singular at the origin, etc.) did not influence much $\rho^2$, while the
form of the kinetic energy (KE) plus mass, whether it is relativistic
($\sqrt{p^2+m^2}$) or non relativistic ($m+p^2/2m$), did change very
sensibly the result. While the relativistic KE gives $\rho^2 = 1.00 \pm
0.05$ for all models that we have considered, the non relativistic KE
provides much larger values, $\rho^2 \simeq 1.3$. Theoretical and
experimental estimates of $\rho^2$ clearly favor the value close to 1,
i.e. the relativistic KE \cite{rho2}. It was a good surprise for us that
the more sensible relativistic KE combined with a covariant way of
boosting the states, the BT method, provides a relatively model
independent and quite reasonable value for $\rho^2$. It was a good
surprise because the relatively stringent lower bound $\rho^2 > 3/4$ \cite{raynal},
valid for any mass operator,  could induce the fear that
a spectroscopically successful mass operator, and its eigenfunctions
which necessarily are very different from the ones saturating the lower
bound, would give a much too large $\rho^2$. This fear could be
reinforced by the fact that gaussian wave functions give a large value
$\rho^2 > 1.2$. It was quite gratifying to realize that the properties
of the spectrum and the short distance potential predicted from
perturbative QCD, plus the relativistic KE was shaping the wave function
towards a lower and better value of $\rho^2$, and towards a larger value
of the wave function at the origin. The latter effect and the related
leptonic decay constants are discussed in \cite{sr2}.

To summarize, we have a model which has several welcome properties, and
provides a very reasonable value of $\rho^2$, and which, among its
beauties, satisfies the Bjorken sum rule \cite{duality},
\cite{dstar} which, in particular, relates the value of $\rho^2$ to the functions which
enter in the $B\to D^{\ast\ast} l \nu$ decay, eq. (\ref{bj}).

This is not a trivial achievement. It comes in this class of models
because the boost of the wave functions is a unitary transformation that
keeps the closure property of the Hilbert space in all frames.  It is
then more than tempting to {\it make predictions with this model for
$B\to D^{\ast\ast} l \nu$} precisely at a time when these branching
ratios start being measured.

In \cite{dstar} we have derived the general formulae giving the HQET
form factors $\tau_{1/2}(w)$ and $\tau_{3/2}(w)$, which are relevant for
$B\to D^{\ast\ast} l \nu$, from the eigenfunctions of the mass operator.

 The content of the BT method has been explained in some detail in
\cite{raynal} and \cite{dstar}. It will not be repeated here. In this
paper we will use the same set of mass operators fitted to the experimental
meson spectrum which has proven to provide a reasonable value of
$\rho^2$ \cite{rho2}. In section 2 we  describe the four models from
which we borrow the mass operator, and our diagonalisation procedure. In
section 3 we  derive the formulae for the partial widths. In section
4 we 
check the Bjorken sum rule. In section 5 we describe the shape
of the functions $\xi(w)$ and $\tau_j(w)$.  In section 6 we will present our predictions
for semileptonic branching ratios and compare them with experiment and
other models. Finally we conclude.

\section{Diagonalization of the mass operator}\label{sec-cov}

\subsection{Mass operators}

The method of Bakamjian-Thomas \cite{BT} \cite{mathiot} provides a fully covariant
\cite{raynal} description of the current matrix elements in the infinite
mass limit, once given a mass operator, i.e. a Hamiltonian operator in
the meson rest frame. For the latter the only condition that we impose
is that it has to describe correctly the meson spectrum.  We did not try
to invent our own Hamiltonian but rather resorted to literature. We have
chosen four quark models: Godfrey and Isgur (GI) \cite{godfrey}, Veseli
and Dunietz (VD)
\cite{veseli}, Cea, Colangelo, Cosmai and Nardulli (CCCN) 
\cite{cea} and finally  Isgur, Scora, Grinstein and Wise (ISGW) \cite{scora}. 
The GI model is rather complex and because of that, one may not agree on
all the ingredients or assumptions which enter it, or one may find it
difficult to evaluate their respective impact on the final result.
Nevertheless, what makes the model outstanding is its covering of the
whole spectroscopy, from light to heavy quarks. The GI model includes
relativistic features, among which the square root kinetic energy:
\beq 
K = \sqrt{\vec{p}_1^{\ 2} + m_1^2} + \sqrt{\vec{p}_2^{\ 2} + m_2^2} \ \
\ ,
\label{kinrel} 
\end{equation}
Starting from the linear plus Coulomb potential complemented with
spin-spin and spin-orbit forces, GI apply a complicated procedure to
smooth the singular parts of the potential (using for example an
$\alpha_s(Q^2)$ which is finite when $Q^2 \to 0$) and smearing the
potential via momentum dependent terms meant to mimic relativistic
effects. Since we consider in this paper the infinite mass limit $m_1
\to \infty$, we have thrown away in the present calculations all terms
which vanish in that limit, except when checking our numerical code
against their spectrum. We use the following set of parameters in the
notations of \cite{godfrey}:
\beq
b= 0.18\, \mbox{GeV}^2,\, c=-253 \,\mbox{MeV},\, \sigma_0= 1.8
\,\mbox{GeV},\, s=1.55,\, \epsilon_{SO}(V)=-.035,\,
\epsilon_{SO}(S)=.055 \end{equation} $$
\alpha_s(Q^2)= 0.25 e^{-Q^2} + 0.15 e^{-Q^2/10}+ 0.20e^{-Q^2/1000},\,\,
 \gamma_1=0.5,\, \gamma_2=\frac{\sqrt{10}}2,\, \gamma_3=\frac{\sqrt{1000}}2$$ 
$$ m_1=10^4\,\mbox{GeV},\, \,m_2=0.22 \,\mbox{GeV}.  $$ where the mass
$m_1=10^4$ GeV is taken as a good approximation of $m_1=\infty$
\footnote{When computing the $B$ meson decay widths, we will use the
amplitudes computed at $m_1= 10^4$ GeV but the phase space factors will
be taken with the physical masses of the $B$, the $D^{(\ast)}$ and
$D^{\ast\ast}$.}.

The VD model is a genuine linear plus coulomb model fitted to the beauty
and charmed flavored mesons:
\beq H= K -\frac 4 3 \frac{\alpha_s} r + br + c \end{equation}
where $K$ is the relativistic kinetic term (\ref{kinrel}) and we use
\beq \alpha_s = 0.498,\, b= 0.142 \,\mbox{GeV}^2,\, c=-350 \,\mbox{MeV},\, 
m_1=10^4\,\mbox{GeV},\, m_2=300 \,\mbox{MeV}\end{equation}

The CCCN model \cite{cea} uses a potential \`a la Richardson
\cite{richardson}
\beq H= K + \frac{8\pi}{33-2n_f} \Lambda\left( {\Lambda r 
-\frac{f(\Lambda r)}{\Lambda r}}\right)\end{equation}
with $K$ in (\ref{kinrel}) and
\beq f(t)= \frac 4 \pi \int_0^\infty  dq  \frac {\sin(qt)}q \left (
\frac 1 {\ln (1+q^2)} - \frac 1 {q^2}\right).\end{equation}
Richardson's potential has the property to be linear for large $r$ and
to behave as predicted by asymptotic freedom for $r\to 0$:
\beq
V(r) {\longrightarrow}
\frac {8\pi}{33-2n_f} \frac 1 {r \ln (\Lambda r)}.
\end{equation}
We use the set of parameters:
\beq
\Lambda = 0.397 \,\mbox{MeV},\,\, n_f=3,\, m_1=10^4\,\mbox{GeV},\,\,
  m_2=38 \,\mbox{MeV}  
\end{equation}
and, as in \cite{cea}, we truncate the potential to a constant value,
\beq V(r)=V(r_m),\, \mbox{for}\, r\le r_m,\, \mbox{with}\,\quad r_m=
 \frac {0.8 \pi}{5.33} \mbox{GeV}^{-1}\end{equation}

Finally, the ISGW model \cite{scora}, contrary to the three previous
ones, has a non relativistic kinetic energy term
\beq H=\frac {p_1^2}{2 m_1}+\frac {p_2^2}{2 m_2} - \frac {4 \alpha_s}{3
r}
 + br + c\end{equation}
with the set of parameters
\beq \alpha_s=0.5,\,\, b= 0.18 \mbox{GeV}^2,\,\, c= -840 \mbox{MeV},\,\, 
 m_1=10^4\mbox{GeV},\,\, m_2=330 \mbox{MeV}\end{equation}. 

In reference \cite{scora} form factors are computed according to the non-relativistic dynamics, with some adjustment of the parameters. We insist that in this paper we do not use the latter estimate, but 
we only take from \cite{scora} the Hamiltonian at rest  and compute the form factors 
according to BT method. The result is indeed totally different. The label ISGW refers to BT estimate of form factors from ISGW Hamiltonian, while, when necessary, we will refer to ISGW's computation of form factors as ISGW-nonrel.

\subsection{Diagonalization}

For each of the four mass operators we have diagonalized the sectors
$l=0$ and $l=1$ using a basis of harmonic oscillator eigenfunctions
which is truncated by keeping only the $n_{max}+1$ lowest states. The code
first computes the matrix elements of the Hamiltonian in this truncated
Hilbert space, and then diagonalizes the matrix keeping the eigenvectors
and the eigenvalues.  We have performed all the calculations with
$n_{max}=10, 15, 20$ and for the GI model with $n_{max}=24$. Before
sending the mass of the heavy quark to $10^4$ GeV, we have checked,
using the masses $m_b, m_c$ fitted by the authors, that we had good
agreement with them for the meson masses. In the case of GI the
check needed the introduction of spin dependent terms which disappear in
the infinite mass limit. VD has singular eigenfunctions for $r\to 0$ due
to the coulombic singularity. CCCN cut the coulomb-logarithmic
singularity at $r_m$, but a discontinuity results in the potential at
$r_m$. Therefore, VD's eigenvalues in the $l=0$ sector converge rather
slowly when $n_{max}\to\infty$, and this is also the case to a smaller
extent for CCCN. This does not prevent a rapid convergence of the matrix
elements which we will consider in this paper. Only the leptonic decay
constants are infinite and the sum rules of \cite{sr1} are ill-defined
for the VD potential \cite{sr2}.

\section{Analytic formulae for $\boldsymbol{B \to D^{\ast\ast}}$ decay} 
\subsection{$\boldsymbol{\xi(w)}$, $\boldsymbol{\tau_{1/2}(w)}$ and $\boldsymbol{\tau_{3/2}(w)}$}

It is well known, \cite{IW}, that in the heavy quark limit the
current matrix elements between $B$ and the even parity $D^{\ast\ast}$
states (odd parity $D^{(\ast)}$) can be expressed as functions of two (one) universal functions,
$\tau_{j}(w)$ ($\xi(w)$), where $w\equiv v\cdot v'$ and where $j=1/2, 3/2$ 
is the total
angular momentum carried by the light quantas (light quarks and gluons)
in the final states. $j$ is a conserved quantum number in the infinite
mass limit. In \cite{dstar} (\cite{raynal}), we have derived the expressions for
$\tau_{\scriptscriptstyle{1/2}}(w)$ and $\tau_{\scriptscriptstyle{3/2}}(w)$ ($\xi(w)$) in the BT type of models, once given the eigenfunctions of the mass
operator:

\begin{equation}
		{\xi(w)  =  \frac1{v'.v+1}  \int \frac{d\vec p_2}{(2\pi)^3}  
\frac{\sqrt{(p_2.v')(p_2.v)}}{p^0_2}}
				\frac{p_2.(v'+v) + m_2(v'.v+1)}{\sqrt{(p_2.v'+m_2)(p_2.v+m_2)}}\; 
\varphi'((p_2.v')^2 - m_2^2)^{\displaystyle*}  \varphi((p_2.v)^2 - m_2^2) 
\end{equation}

\begin{multline}
\tau_{\scriptscriptstyle{1/2}}(w)\ =\ \dfrac{1}{2\sqrt 3}\,\int\,
\frac{d\vec p_2}{(2\pi)^3}\,\frac{1}{p^o_2}\,
\frac{\sqrt{(p_2\cdot v')(p_2\cdot v)}}{\sqrt{(p_2\cdot v'+m_2)
(p_2\cdot v+m_2)}}
\phi_{\frac12}((p_2.v')^2 - m_2^2)^{\displaystyle*}
\varphi((p_2.v)^2 - m_2^2)\\
\times\,\frac{(p_2\cdot v)(p_2\cdot {v'}+m_2)\ -\ (p_2\cdot {v'})
(p_2\cdot {v'}+(v\cdot {v'})m_2)
\ +\ (1-v\cdot {v'})m_2^2}{1-v\cdot {v'}}
\end{multline}
and
\begin{multline}
\tau_{\scriptscriptstyle{3/2}}(w)\ =\ 
\dfrac{1}{\sqrt 3}\,\dfrac{1}{1-(v\cdot {v'})^2}\,
\int\,\frac{d\vec p_2}{(2\pi)^3}\,\frac{1}{p^o_2}\,
\frac{\sqrt{(p_2\cdot v')(p_2\cdot v)}}{\sqrt{(p_2\cdot v'+m_2)
(p_2\cdot v+m_2)}}
\phi_{\frac32}((p_2.v')^2 - m_2^2)^{\displaystyle*}
\varphi((p_2.v)^2 - m_2^2)\\
\times\ \left\{ {\dfrac32\,\frac{1}{1+v\cdot {v'}}\,\left[ {\,p_2\cdot 
(v+{v'})}\right]^2\ -\ 
(p_2\cdot v)(2\,p_2\cdot {v'}-m_2)\ -\ (p_2\cdot {v'})\left[ {\,p_2\cdot
{v'} +(v\cdot {v'})m_2}\right]\ -\ \dfrac{1-v\cdot {v'}}{2}\,m_2^2
}\right\}
\end{multline}
where $\varphi(\vec p^{\,2})$ ($\phi_{j}(\vec p^{\,2})$) is the radial
part of the wave function of the $B$ ($D^{\ast\ast}$'s). A detailed
account of our conventions can be found in \cite{dstar}.

\subsection{Differential decay widths}

Leaving aside radial excitations, there are four $l=1$ states, the $D^{\ast\ast}$, that we will
label according to the common use: $D^{\ast}_0$, $D^{\ast}_{1}$ with
$j=1/2$ and $D_1$, $D^{\ast}_{2}$ with $j=3/2$. The lower index stands
for the total angular momentum $J$.  Two experimental resonances, the
$D_1(2420)$ and the $D^\ast_2(2460)$, are considered as good candidates
for the $j=3/2$ states. In the infinite mass limit, to which we stick
here, the mixing between both $J=1$ states ($j=1/2$ and $j=3/2$) is
forbidden. The $O(1/m_c)$ mixing seems to be small \cite{falk}.

To make short a long story, we have checked the formulae in \cite{IW}
with which we agree except for three misprints: in eq.(45) in \cite{IW}
one should read in the denominator $32\pi^3$ instead of $32\pi^2$ and
$r\equiv m_{X_{Q_j}}/m_{P_{Q_i}}$ (for example $r\equiv
m_{D^{\ast\ast}}/m_B$) instead of $r\equiv m^2_{X_{Q_j}}/m^2_{P_{Q_i}}$.
Finally we agree with table III in \cite{IW} if one defines
$\delta\equiv [2rw^2 -(1+4r +r^2)w+2(1+r^2)]/4r$ instead of
$\delta=[2rw^2 -(1+4r +r^2)w+2(1+r^2)]/2r$.

 While we have computed the wave functions and the universal functions
$\tau_j(w)$ in the infinite mass limit, the kinematics is taken with the
physical masses of the particles (or, in the absence of measurement, some
estimated physical masses), i.e.
\begin{equation}
 m_B=5.279,\quad m_D=1.865,\quad m_{D^\ast}=2.010,\quad
m_{D^{\ast}_0}=2.360,\quad m_{D^\ast_1}=2.420, \quad m_{D_1}=2.422,\quad
m_{D_2^\ast}=2.459,
\label{masses}\end{equation}
We also use
\begin{equation}
V_{cb}=0.041, \qquad G_F=1.166389\, 10^{-5} \mbox{GeV}.
\end{equation}
and $r=m_X/m_B$, $X$ being the charmed final meson.
The calculation of the differential decay widths is then standard, leading to
(taking the mass of the final lepton to zero)
\begin{eqnarray}
\frac {d \Gamma}{dw}(B\to D l \nu) &=& \vert V_{cb}\vert^2 \frac
 {G_F^2 m_B^5}{48 \pi^3}
r^3(w+1)(w^2-1)^{1/2} (w-1)(1+r)^2 \vert \xi(w)\vert^2\label{BtoD}
\\
\frac {d \Gamma}{dw}(B\to D^\ast l \nu) &=& \vert V_{cb}\vert^2 \frac 
{G_F^2 m_B^5}{48 \pi^3}
r^3(w+1)(w^2-1)^{1/2}[(w+1)(1-r)^2+4w(1+r^2-2rw)]\vert
\xi(w)\vert^2\label{BtoDs}
\\
\frac {d \Gamma}{dw}(B\to D^{\ast}_0 l \nu) &=&
 \vert V_{cb}\vert^2 \frac {G_F^2 m_B^5}{48 \pi^3}
4r^3(w-1)(w^2-1)^{1/2} (w+1)(1-r)^2\vert
\tau_{1/2}(w)\vert^2\label{BtoD0}
\\
\frac {d \Gamma}{dw}(B\to D_1^\ast l \nu) &=&
 \vert V_{cb}\vert^2 \frac {G_F^2 m_B^5}{48 \pi^3}
4r^3(w-1)(w^2-1)^{1/2}[(w-1)(1+r)^2+4w(1+r^2-2rw)]
\vert \tau_{1/2}(w)\vert^2\label{BtoDs1}\\
\frac {d \Gamma}{dw}(B\to D_1 l \nu) &=& \vert V_{cb}\vert^2 
\frac {G_F^2 m_B^5}{48 \pi^3}
2r^3(w+1)(w^2-1)^{3/2} [(w-1)(1+r)^2+w(1+r^2-2rw)] \vert
\tau_{3/2}(w)\vert^2\label{BtoD1}\\
\frac {d \Gamma}{dw}(B\to D_2^\ast l \nu) &=& \vert V_{cb}\vert^2 
\frac {G_F^2 m_B^5}{48 \pi^3}
2r^3(w+1)(w^2-1)^{3/2} [(w+1)(1-r)^2+3w(1+r^2-2rw)] \vert
\tau_{3/2}(w)\vert^2\label{BtoDs2}
\end{eqnarray}

We have also computed $d\Gamma/d E_l$ where $E_l$ is the energy of the
final lepton. We do not think it very useful to write down the lengthy
resulting formulae, we will show some plots later on.

\section{Bjorken sum rule}

It has been demonstrated in \cite{duality} and \cite{dstar} that the
heavy quark models \`a la Bakamjian-Thomas satisfy exactly the Bjorken sum
rule \cite{bjorken},
\cite{IW} thanks to the closure property of the Hilbert space. 
In this section, we would like to make this statement concrete, while
 providing a flavor of the value and shape of the $\tau_j$ functions. 

In table \ref{tab:tau}, we check the derivative Bjorken sum rule, obtained by a derivation of both sides of full Bjorken sum rule at $w=1$:
\begin{equation}
 \rho^2 = \sum_{k} \left ( \left | \tau_{1/2}^{(k)}(1) 
\right |^2 + 2 \left |
\tau_{3/2}^{(k)}(1) \right |^2 \right ) + \frac14. \label{bj}
\end{equation}
We have given the values of $\tau_j(w)$ for $w=1.0$ and $w=1.5$ to give
some feeling of the decrease of $\tau_j$ with $w$.  Adding only the
ground state $l=1$ mesons in the r.h.s gives a contribution
about 20 \% below the l.h.s, $\rho^2$ (35 \% in the case of the linear + Coulomb
potential of VD). We have computed the contributions from the first
radial excitation of the $l=1$ mesons. Now the discrepancy is of a few
percent except for the VD, still about 20 \% too low. Finally, adding
all the $l=1$ eigentstates in our truncated Hilbert space, we get a 1 \%
agreement with the directly computed $\rho^2$, {\it thus confirming that
the models \`a la BT do incorporate naturally the Bjorken sum rule}.

The larger value of $\rho^2$ in ISGW model is related, via the derivative Bjorken
sum rule, to both a larger $\tau_{1/2}(1)$ and a larger $\tau_{3/2}(1)$.

\vskip 5mm
\begin{table}[htb]
\begin{center}
\begin{tabular}{|c|c|c|c|c|}\hline
 & GI & VD & CCCN & ISGW \\ \hline $\tau_{1/2}(1)$ & 0.2248(1) & 0.13(1)
& 0.059(1) & 0.3424(1) \\ $\tau_{3/2}(1)$ & 0.5394(1) & 0.43(2) &
0.515(1) & 0.5918(2) \\ $\tau_{1/2}(1.5)$ & 0.1151(1) & 0.10(1) &
0.042(2) & 0.2116(2) \\ $\tau_{3/2}(1.5)$ & 0.2767(1) & 0.23(1) &
0.270(1) & 0.2700(1) \\ $\vert\tau_{1/2}(1)\vert^2$ & 0.051 & 0.017 & 0.004 &
0.117 \\ $2\vert\tau_{3/2}(1)\vert^2$ & 0.582 & 0.37 & 0.531 & 0.701 \\ ground
state BSR & 0.882 & 0.64 & 0.79 & 1.068 \\ $\tau_{1/2}^{(1)}(1)$ &
0.1965(3) & 0.15(1) & 0.249(2) & 0.2290(2) \\ $\tau_{3/2}^{(1)}(1)$ &
0.2147(1) & 0.25(2) & 0.216(1) & 0.2522(4) \\
$\vert\tau_{1/2}^{(1)}(1)\vert^2+2\vert\tau_{3/2}^{(1)}(1)\vert^2$ & 0.130 & 0.15 & 0.16 &
0.18 \\ BSR & 1.023 & 0.97 & 0.97 & 1.283 \\ $\rho^2$ & 1.023(3) &
0.98(2) & 0.97(2) & 1.283(1) \\
\hline
\end{tabular}
\end{center}
\caption{\leftskip 1pc \rightskip 1pc
\baselineskip=10pt plus 2pt minus 0pt {\sl {
Results for $\tau_j(1)$ and $\tau_j(1.5)$ form factors into the  $D^{\ast \ast}$'s (lowest
 l=1 states). The errors in parenthesis is for
truncation (estimated by varying $n_{max}$). The line ``ground state
BSR'' indicates the r.h.s of the derivative Bjorken sum rule (\ref{bj}) in which the $\tau$'s have only
been taken for the lowest orbitally excited final states.  Adding only
the contribution of the first radially excited l=1 final states (indicated by the superindex (1)) almost fills the gap with
$\rho^2$. The line ``BSR'' gives the contribution to the r.h.s of the derivative Bjorken sum
rule from all the states in our diagonalization procedure. The agreement
with the l.h.s., $\rho^2$, is almost perfect.}}}

\label{tab:tau}\vskip 5mm
\end{table}

The full Bjorken sum rule writes \cite{IW}
\begin{equation}
1=\frac {w+1}2 |\xi(w)|^2+(w-1)\left[\sum_{n=1}^{n_{max}}\frac{w^2-1}2
|\xi^{(n)}(w)|^2 + 2 \sum_{n=1}^{n_{max}}|\tau_{1/2}^{(n)}(w)|^2 +(w+1)^2
\sum_{n=1}^{n_{max}}|\tau_{3/2}^{(n)}(w)|^2\right]+ .....\label{fbj}
\end{equation}
where the dots represent the $l>1$ states. In \cite{IW}, $n_{max}$
corresponds to some scale dependent cut-off, which does not exist in our
case since we did not introduce QCD radiative corrections. We will use
for $n_{max}$ the higher state in our truncated Hilbert space. Our
results are exhibited in table \ref{tab:bj}.

\vskip 5mm
\begin{table}[htb]
\begin{center}
\begin{tabular}{|c|c|c|c|c|}\hline
 & GI & VD & CCCN & ISGW \\ \hline w=1.0 &1.000 &1.000 &1.000 &1.000 \\
w=1.1 &0.9864(2) &0.988(2)&0.988(2)&0.9793(1)\\ w=1.2
&0.9546(5)&0.962(3)&0.959(3)&0.9337(1)\\ w=1.3
&0.913(1)&0.928(5)&0.921(5)&0.8781(1)\\ w=1.4
&0.870(2)&0.890(6)&0.879(2)&0.8201(1)\\ w=1.5
&0.825(3)&0.85(1)&0.83(1)&0.7633(1)\\
\hline
\end{tabular}
\end{center}
\caption{\leftskip 1pc \rightskip 1pc
\baselineskip=10pt plus 2pt minus 0pt {\sl {
r.h.s of the full Bjorken sum rule (\ref{fbj}) as a function of $w$ adding all the $l=0$ and
$l=1$ states. The error is for the truncation.} }}
\label{tab:bj}\vskip 5mm
\end{table}

 For $w=1$, the sum rule is a triviality from the normalization
$\xi(1)=1$ and the vanishing of other contributions. The larger $w$, the
more room is left for $l>1$ states. However, even for $w=1.5$, which is
beyond the end of the physical domain ($w_{max}\simeq 1.32$ for
$D^{\ast\ast}$'s), less than 20 \% is due to $l>1$ except for the
non-relativistic Hamiltonian. Again, as for $\rho^2$, we note a general
agreement
 between the three models using the relativistic kinetic energy 
(\ref{kinrel}), and a sensibly larger ($\sim$ 25 \%) need of 
$l>1$ states  for
the ISGW model which uses a non-relativistic kinetic energy.

\begin{figure}[p]
\epsfysize=100mm
\hspace{20mm}\epsfbox{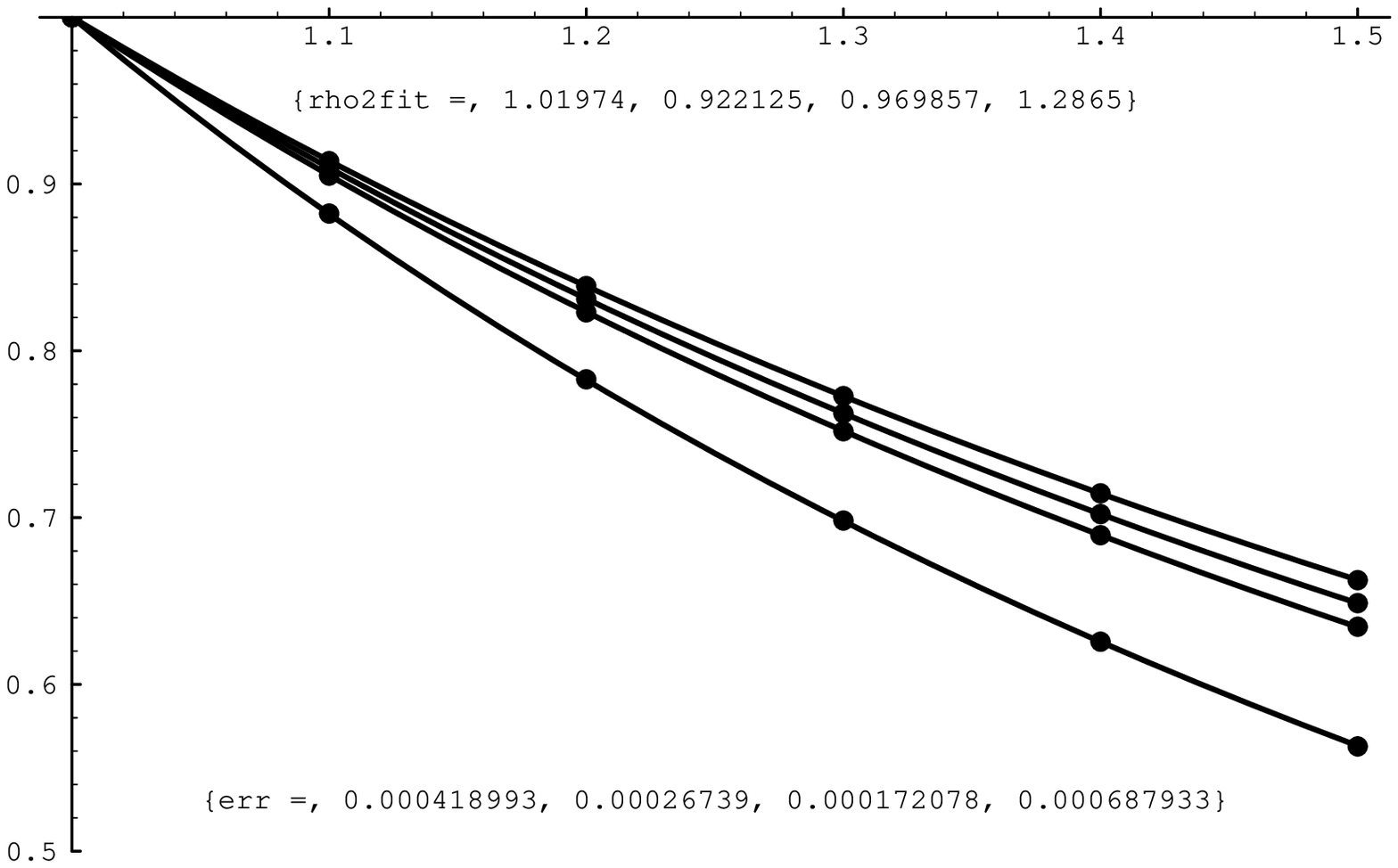}
\caption{\leftskip 1pc \rightskip 1pc
\baselineskip=10pt plus 2pt minus 0pt {\sl 
{Plots of $\xi(w)$. The dots are $\xi(w)$ calculated for $w=1.0 - 1.5$
The curves are functions $(2/(1+w))^{2 \rho^2_{fit}}$ with
$\rho^2_{fit}$ fitted to the above-mentioned points. From upper to lower
the curves correspond
to VD, CCCN, GI and ISGW models. The numbers indicate $\rho^2_{fit}$
 for GI, VD, CCCN and ISGW in this order.}
}}
\label{fig:fig1}
\vskip -.3cm\end{figure}

\section{Numerical predictions for the functions $\boldsymbol{\xi(w)}$ and $\boldsymbol{\tau_j(w)}$.}

\subsection{The difference ${\boldsymbol{\tau_{3/2}(1)-\tau_{1/2}(1)}}$}

Table \ref{tab:tau} clearly shows a dominance of $\tau_{3/2}(w)$ over
$\tau_{1/2}(w)$ ($w=1-1.5,$) of more than a factor 2,  except for the non-relativistic
ISGW model. This feature
 is at the origin of a predicted suppression of the semi-leptonic
decay into $D^\ast_1(1/2) l \nu$ and $D^\ast_0 l \nu$. It should be noted
that this $\tau_{3/2}(w)\gg \tau_{1/2}(w)$ is only valid for the
groundstate $l=1$ states. Both $\tau$'s are of the same order for the
first radially excited $l=1$ states.

The  difference between $\tau_{3/2}$ and $\tau_{1/2}$ is not due to the difference between the 1/2 and 3/2 internal wave functions at rest coming from the spin-orbit force, which is small and
has a rather moderate effect. It is essentially
due to the relativistic structure of the matrix
element in terms of these wave functions. One can gain understanding of this effect at least  near $w=1$ by returning to the analysis made in 
\cite{duality}, in a frame where $\vec{v}$ and $\vec{v'}$ are small. There, the current
has been shown (eq.(30) in \cite{duality}) to have one spin-independent piece and one which is spin-dependent.
The latter, corresponding to the {\it light spectator quark} Wigner rotations, is responsible for the large
 difference between $\tau_{3/2}(1)$ and $\tau_{1/2}(1)$ :
\begin{equation} \tau_{3/2}(1) -\tau_{1/2}(1) \simeq {\frac 1  {2\sqrt{3}}}\left\vert \int dp\,p^2\,{p\,\phi^\dagger(p)} \,
\frac p  {p_0 +m}\, \varphi(p)\,\,\right\vert ,
\end{equation}
where $\phi$ is $\phi_{1/2} \simeq \phi_{3/2}$ . In a non relativistic expansion, this Wigner rotation effect is small, of relative order ${\cal O} (v^2/c^2)$ with respect to the main, spin-independent term. That it may be large in spite of this illustrates the fact that the system is ultrarelativistic in the GI case, while it is less relativistic in the ISGW case. For $\rho^2$, the
Wigner rotation effect was found rather small, partly due to a small algebraic coefficient, and partly to the fact that in $\rho^2$ it is squared.

\subsection{Shape of the functions $\boldsymbol{\xi(w)}$ and $\boldsymbol{\tau_j(w)}$}

\begin{figure}[p]
$$\begin{array} {cl}\epsfbox{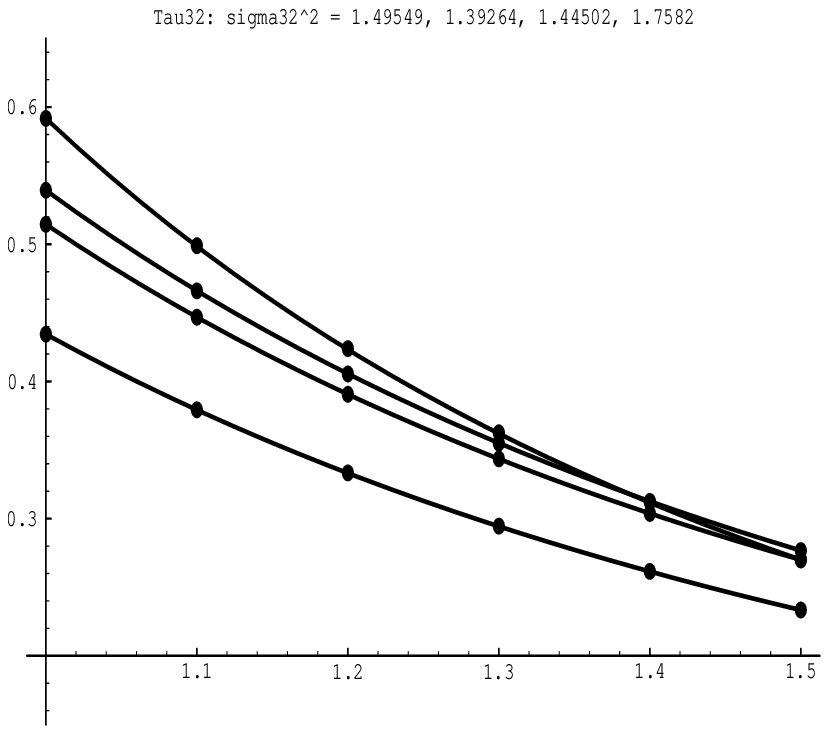} & \epsfbox{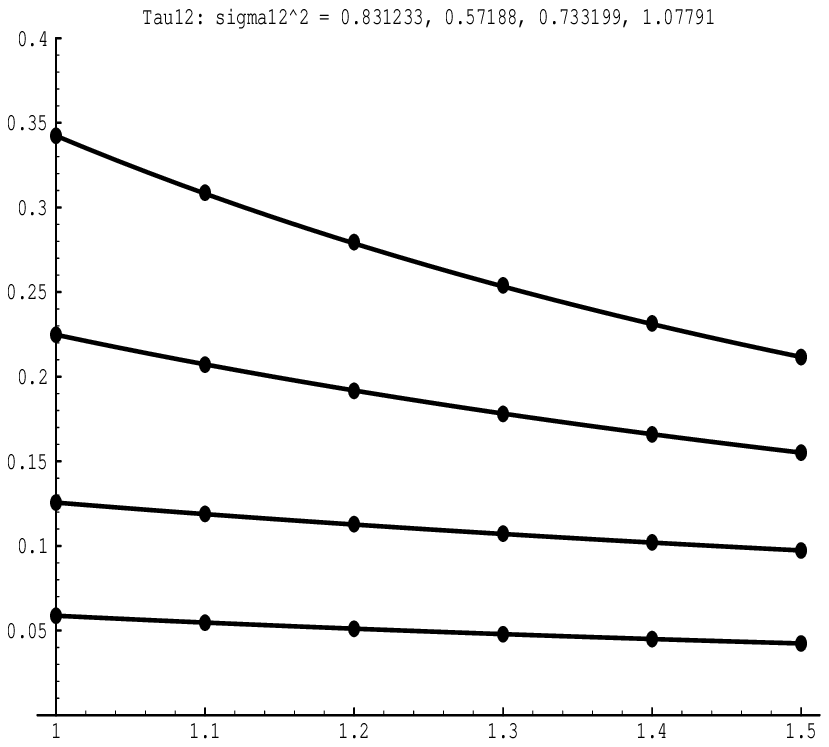} \end{array}$$
\caption{\leftskip 1pc \rightskip 1pc
\baselineskip=10pt plus 2pt minus 0pt {\sl {
Plots of $\tau_{3/2}(w)$ to the left and $\tau_{1/2}(w)$ to the right. 
The dots are $\tau_j(w)$ calculated for $w=1.0 - 1.5$
The curves are functions $\tau_j(1)(2/(1+w))^{2 \sigma^2_{j}}$ with 
$\sigma^2_{j}$
fitted to the above-mentioned points. From upper to lower the curves
 correspond 
to ISGW, GI, CCCN, VD (ISGW, GI, VD, CCCN) for $\tau_{3/2}$
($\tau_{1/2}$).
 The numbers indicate $\sigma^2_{j}$ for GI, VD, CCCN and ISGW in this order.
}}}
\label{fig:fig2}
\vskip -.3cm\end{figure}

In order to analyze the shape of the function $\xi(w)$, we have
fitted the predictions of our models to several standard one parameter analytic formulae \cite{neubert}. 
\begin{eqnarray} 
\xi_{\text{NR}}(w)&=&\frac 2 {w+1} \exp\left[-(2 \rho^2-1)\frac{w-1}{w+1}\right],\label{NR}\\
\xi_{\text{ISGW-nonrel}}(w)&=&\exp\left[-\rho^2(w-1)\right],\label{IS}\\
\xi_{\text{pole}}(w)&=&\left(\frac 2 {1+w}\right)^{2 \rho^2_{fit}},\label{pole}\\
\xi_{\text{lin}}(w)&=&1-\rho^2(w-1).
\label{xifit}
\end{eqnarray}
 $\xi_{\text{NR}}(w)$ has been proposed in \cite{nr}, $\xi_{\text{pole}}(w)$ in \cite{neubert} and $\xi_{\text{ISGW-nonrel}}(w)$ in \cite{scora}.
It should be repeated that $\xi_{\text{ISGW-nonrel}}(w)$ is totally different 
from what is 
referred to as the ISGW model throughout this paper. We only take from the ISGW model the mass operator, and compute $\xi(w)$ consistently in the framework of the BT method. 
As a result, as we shall see, our $\xi(w)$ deduced from the ISGW model is much better fitted by
$\xi_{\text{pole}}(w)$ than by $\xi_{\text{ISGW-nonrel}}(w)$. We have labelled the latter by ISGW-nonrel in order to avoid any confusion. 

We fit $\rho^2$ respectively to these four analytic shapes and estimate the quality
of the fit by computing:
\begin{equation}
\delta=\left\{ \sum_{w_i=1-1.5} \left[\xi(w_i)-\xi_{fit}(w_i)\right]^2\right\}^{1/2}  
\end{equation}
where $\xi_{fit}$ is one of the functions in (\ref{NR})-(\ref{xifit}) and where $w_i$ runs
from 1 to 1.5 by steps of 0.1. The results are
\begin{equation}
\delta_{\text{NR}}\simeq (5 \text{ to } 9) \, 10^{-3},\,\,\,\delta_{\text{ISGW-nonrel}}\simeq (11\text{ to }14) \, 10^{-3},\,\,\,
\delta_{\text{pole}}\simeq (2 \text{ to } 7) \, 10^{-4},\,\,\,
\delta_{\text{lin}}\simeq (3 \text{ to } 6) \, 10^{-2} \label{delta}
\end{equation} 
$\xi_{\text{pole}}(w)$ wins by far. In fact, in our case, it would deserve the label ``dipole'' since $\rho^2\simeq 1$ and we may claim that the BT models with relativistic kinetic energy predict
\begin{equation}
\xi(w) \simeq\frac {(4 m_B m_{D^{(\ast)}})^2}{(q^2-(m_B+m_{D^{(\ast)}})^2)^2}.
\label{dipole}\end{equation}
although, clearly, this singularity does not correspond to an exchanged particle. This dipole behavior is to be compared to the pole-type behavior which results, for $w\to\infty$, from a na\"\i ve Lorentz contraction, see section 6 in \cite{aleksan}.
Notice that the form factors $A_1(q^2)$ and $f_0(q^2)$ have one ``pole'' less than $\xi(w)$ (eq (17) in \cite{aleksan}), while the other form factors behave like $\xi(w)$. Hence, the result (\ref{dipole}) favors a pole/dipole behaviour 
of the form factors, contrary to the constant/pole advocated in \cite{aleksan}
on the basis of the naive Lorentz contraction. In fact other arguments were also used in \cite{aleksan} and an update of the latter discussion would be welcome in view of the theoretical and experimental progress. 

In \cite{rho2} we have argued that, when using the relativistic kinetic energy, {\it the ground state wave functions are rather close to ground state Coulomb wave function} (see eq. (14) anf figure 1 in \cite{rho2}). It happens that in the limite of a vanishing light quark mass, the ground state {\it Coulomb wave function gives exactly} $\xi(w)=(2/(1+w))^2$, i.e. our best fit $\xi_{\text{pole}}(w)$ with $\rho^2=1$. Clearly the ``dipole'' behaviour of $\xi(w)$ may be traced back to the Coulomb-like shape of the wave function, which in turn stems from the kinetic energy (\ref{kinrel}).
It was also noticed in \cite{rho2} that this coulombic wave function is similar to the NRQCD lattice results. 

The fits with $\xi_{\text{pole}}(w)$ are shown in figure 1. 
The fitted $\rho^2$ are presented in table \ref{tab:fit}. 
 It appears that they  agree very well with the exact slope $\rho^2$ except for the VD
model where a 6\% difference is to be noted. We do not know whether the
latter difference has to do with the strong singularity of the Coulomb
potential in VD model. Needless to say, the other fitting functions in eqs. (\ref{NR}), (\ref{IS}) and (\ref{xifit}) lead to values of $\rho^2_{fit}$ which disagree 
with $\rho^2$ by several percent in the case of $\xi_{\text{NR}}(w)$, up to 30\% in the case of eq. (\ref{xifit}). The latter point has already been discussed in \cite{rho2}.

\begin{table}[htb]
\begin{center}
\begin{tabular}{|c|c|c|c|c|}\hline
 & GI$^\ast$ & VD & CCCN$^\ast$ & ISGW   \\ \hline
$\rho^2_{fit}$        & 1.020 & 0.92 & 0.97 & 1.287  \\
$\rho^2$        & 1.023(3) & 0.98(2) & 0.97(2) & 1.283(1)  \\
$\tau_{1/2}(1)$ & 0.2248(1) & 0.13(1) & 0.059(1) & 0.3424(1)    \\
$\sigma_{1/2}^2$ & 0.83 & 0.57 & 0.73 & 1.08   \\
$\tau_{3/2}(1)$ & 0.5394(1) & 0.43(2) & 0.515(1) & 0.5918(2)  \\
$\sigma_{3/2}^2$ & 1.50 & 1.39 & 1.45 & 1.76   \\
\hline
\end{tabular}
\end{center}
\caption{\leftskip 1pc \rightskip 1pc
\baselineskip=10pt plus 2pt minus 0pt {\sl {
Results of the fit of $\xi(w)$ by $[2/(1+w)]^{2 \rho^2_{fit}}$, and 
of $\tau_j(w)$ by $\tau_j(1)[2/(1+w)]^{2 \sigma^2_{j}}$. For comparison
 the exact $\rho^2$ is repeated. We have starred our preferred models: GI and CCCN which have a relativistic kinetic energy and have a regular potential at $r\to 0$.}
}}
\label{tab:fit}\vskip 5mm
\end{table}
In a recent paper S. Simula \cite{simula} has computed the Isgur-Wise function from the light-front constituent model of \cite{cardarelli} based on the GI Hamiltonian. Not only does he get $\rho^2=1.03$ in the infinite mass limit, but the detailed agreement of his solid line in figure 4 with our results for GI model (figure \ref{fig:fig1}) is very striking. This confirms that the models are indeed equivalent as discussed in \cite{raynal}.
\par Expanding 
\begin{equation}
\xi(w) = 1 -\rho^2(w-1) + c (w-1)^2 +...
\end{equation} 
the different analytic functions in (\ref{NR})-(\ref{xifit}) lead to
\begin{equation}
c_{\text{NR}}=\frac{\rho^4+\rho^2-0.25}2,\quad
c_{\text{ISGW-nonrel}}=\frac {\rho^4}2,\quad
c_{\text{pole}}=\frac{\rho^2(1+2\rho^2)}4.
\end{equation}
For GI (VD) model it leads to
\[
\rho^2=1.02, c_{\text{NR}}=0.90, c_{\text{ISGW-nonrel}}=0.52, c_{\text{pole}}=0.77
\quad\text{(GI)}\]
\begin{equation} \rho^2=0.92,c_{\text{NR}}=0.76, c_{\text{ISGW-nonrel}}=0.42, c_{\text{pole}}=0.65\quad\text{(VD)}
\end{equation}
In the $(\rho^2, c)$ plane, our best fit $(\rho^2, c_{\text{pole}})$ for GI (VD) is slightly above (just on) the ellipse limiting the allowed domains shown in figure 1 of \cite{caprini}. 
  
In figure 2 we also show 
the fits of $\tau_{j}(w)$ according to
\begin{equation}
\tau_j(w)\simeq \tau_j(1)  \left(\frac 2 {1+w}\right)^{2 \sigma^2_{fit}}
\end{equation}

Notice  the stable and rather large slopes $\sigma^2_j \simeq 1.5$ for the dominant $j=3/2$ channel.

\section{Predicted branching ratios}

\subsection{Differential decay widths}

\begin{figure}[p]
\epsfysize=90mm\centerline{\epsfbox{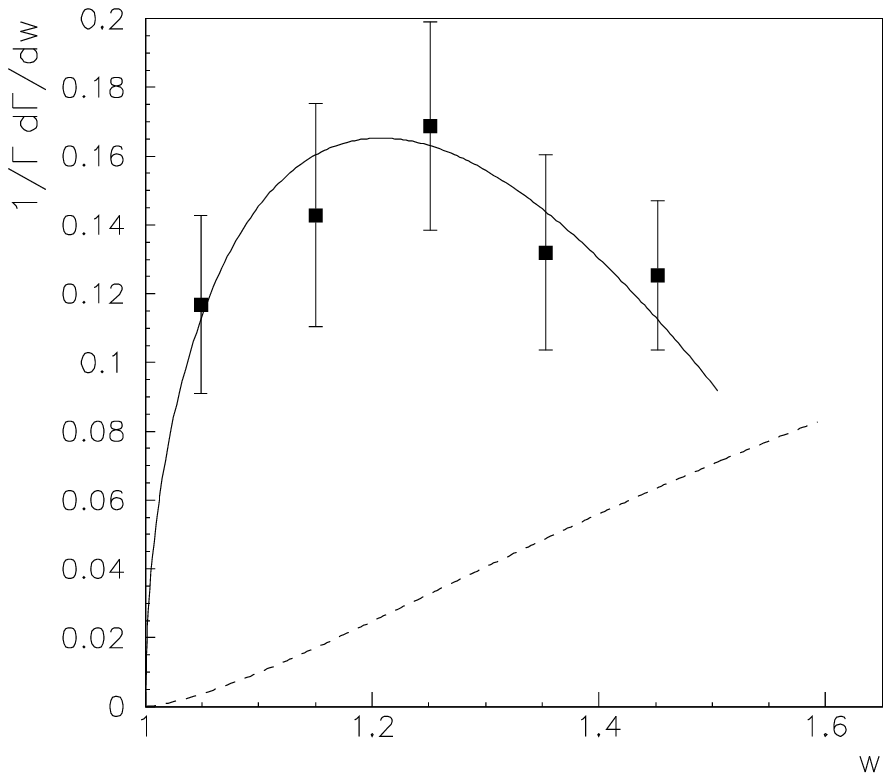}}
\caption{\leftskip 1pc \rightskip 1pc
\baselineskip=10pt plus 2pt minus 0pt {\sl 
{ $\frac{1}{\Gamma}\frac{d\Gamma}{dw}$ of the semileptonic production of
non-orbitally excited $D$, eq. (\ref{BtoD}), (dashed line) and $D^\ast$, eq. (\ref{BtoDs}) (solid line) mesons, compared with  DELPHI measurements for $D^\ast$, normalized to adjust the total number of experimental events to the integral of our theoretical curve. The experimental numbers correspond to bins in $w$ of size 0.1.}}}
\label{fig:fig6}
\vskip -.3cm\end{figure}

We show in figure \ref{fig:fig6} our predictions in the GI model 
for $\dfrac{1}{\Gamma}\dfrac{d\Gamma}{dw}$ of the semileptonic production of
non-orbitally excited $D$  and $D^\ast$ mesons, eqs. (\ref{BtoD}), (\ref{BtoDs}). We compare the latter with DELPHI measurement for $B\to D^\ast$ (fig. 10 in \cite{delphi}) multiplied by a factor such that the total number of good events is adjusted to the integral of our predicted partial width.  The agreement is quite satisfactory.

Moreover, as promised in a preceding section, we also computed the leptonic spectra
$d\Gamma/d E_l$ for each type of $c$-meson produced in the $B$ decays.
There is no analytical formula available, as it is the case for the $d\Gamma/dw$
widths: numerical calculations have to be made.  So we are presenting
these results in the following plots: the dynamics used is, here again, the GI model
and we have only considered the non radially-excited $D, D^\ast, D^\ast_0,
D^\ast_1, D_1 \text{ and }D^\ast_2$ mesons (though it is possible to
compute the same quantities for the radially excited ones).\par

\begin{figure}[p]
$$\begin{array} {cc}\epsfxsize=86mm\epsfbox{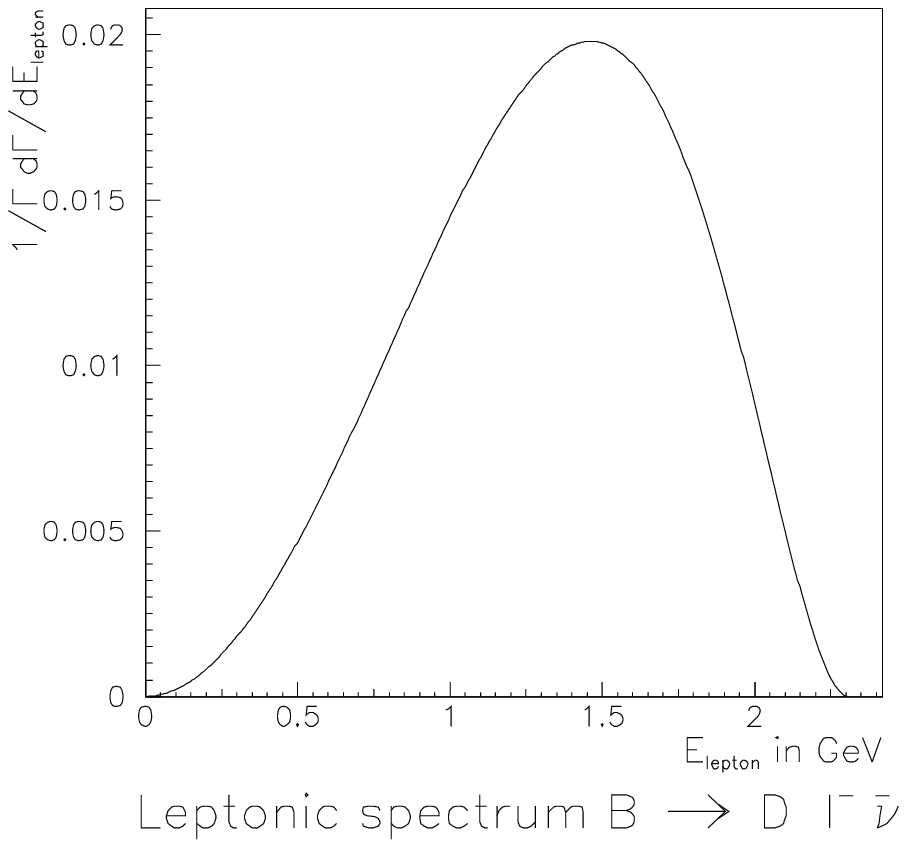} &
\epsfxsize=86mm\epsfbox{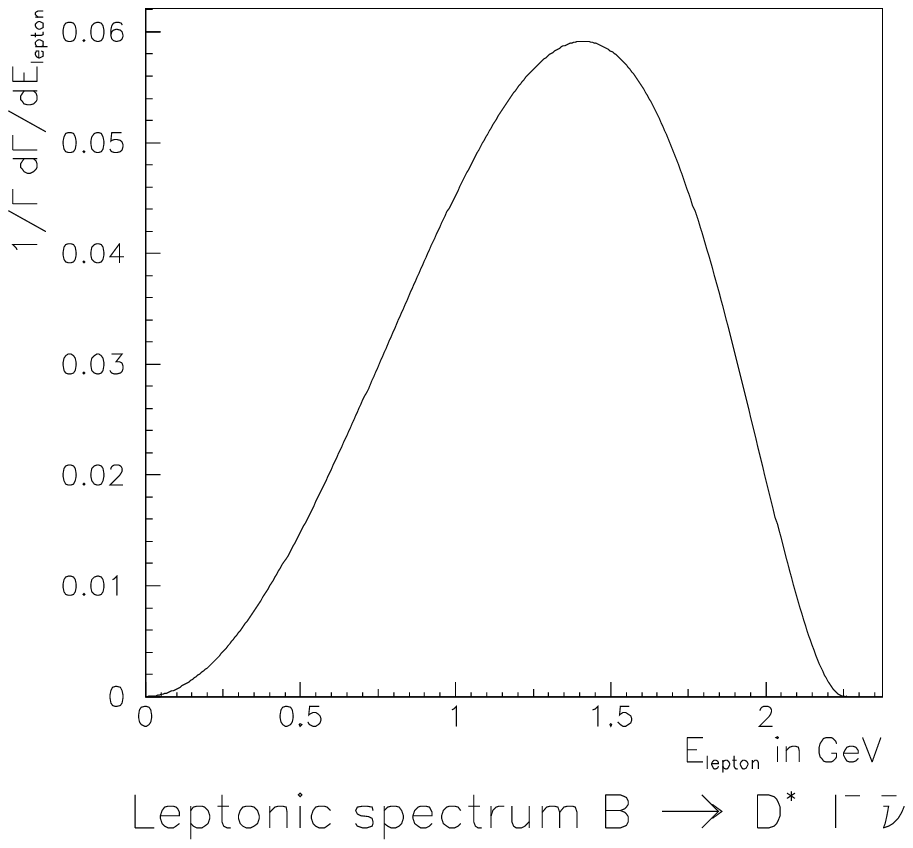}\\
\end{array}$$
\caption{\leftskip 1pc \rightskip 1pc
\baselineskip=10pt plus 2pt minus 0pt {\sl {Leptonic spectra $\dfrac{1}{\Gamma}\dfrac{d\Gamma}{dE_l}$ of the semileptonic production of
$l=0$ $D$ and $D^\ast$ mesons }}}
\label{fig:fig3}
\vskip -.3cm\end{figure}
 
\begin{figure}[p]
$$\begin{array} {cc}
\epsfxsize=86mm\epsfbox{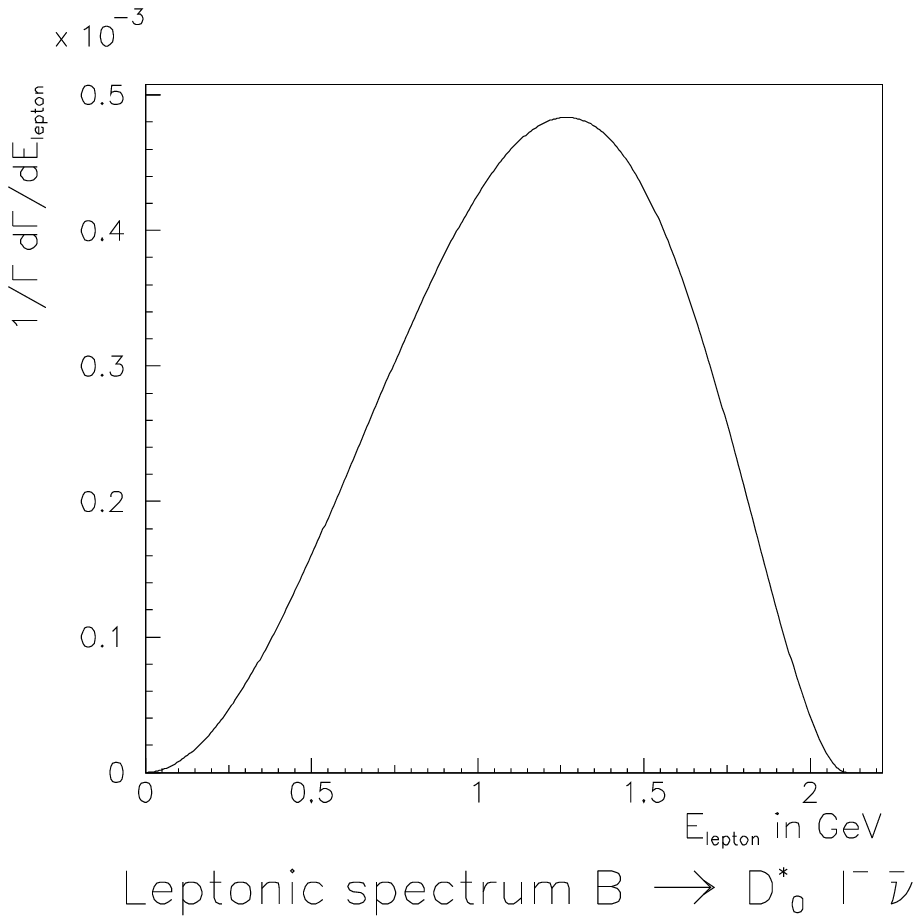} & \epsfxsize=86mm\epsfbox{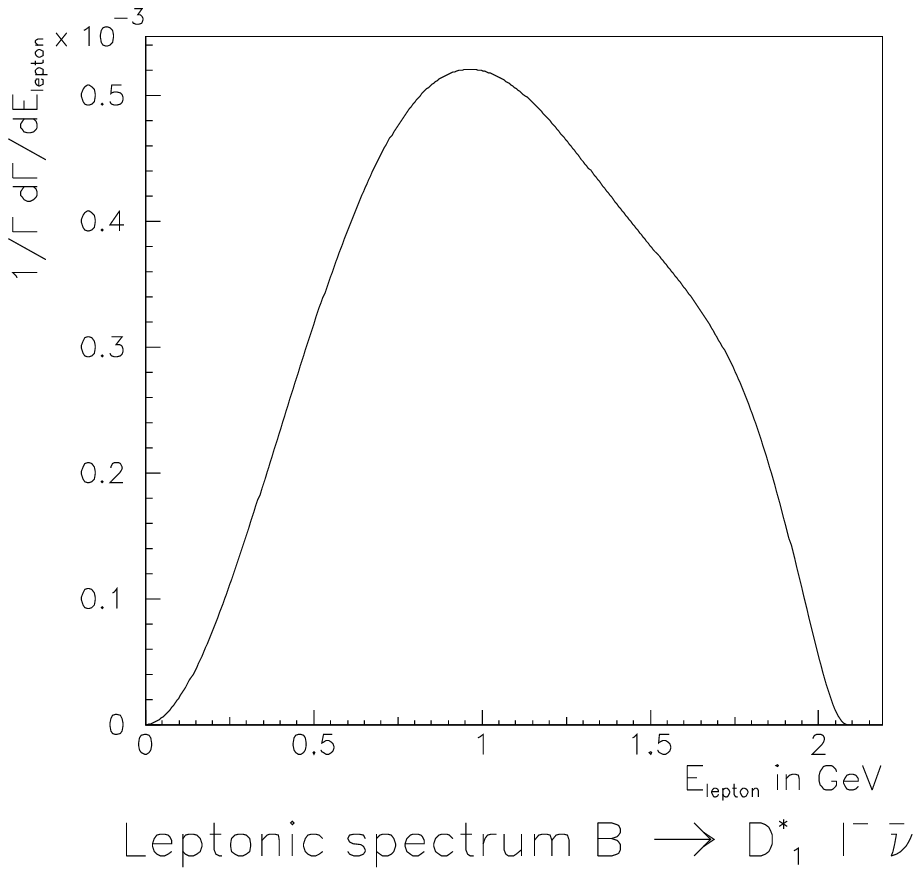}\\
\end{array}$$
\caption{\leftskip 1pc \rightskip 1pc
\baselineskip=10pt plus 2pt minus 0pt {\sl {Leptonic spectra
 $\frac{1}{\Gamma}\frac{d\Gamma}{dE_l}$ of the semileptonic production of
the first $j=1/2$ multiplet of $D^{\ast\ast}$ mesons.}}}
\label{fig:fig4}
\vskip -.3cm\end{figure}
 
\begin{figure}[p]
$$\begin{array} {cc} 
\epsfxsize=86mm\epsfbox{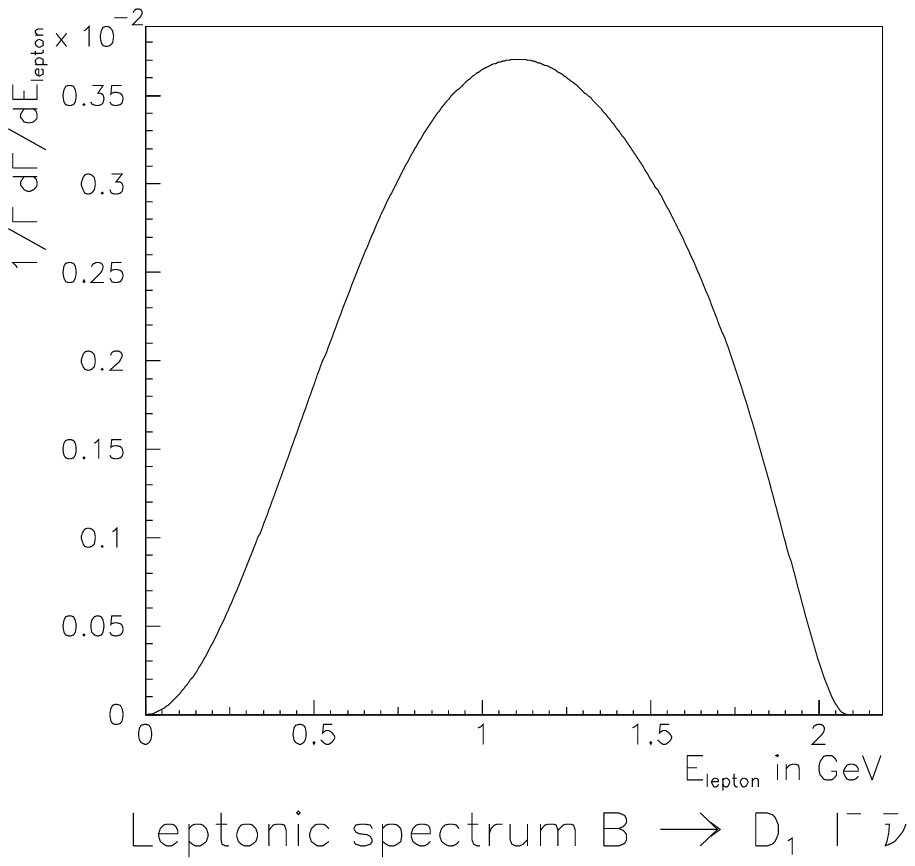} & \epsfxsize=86mm\epsfbox{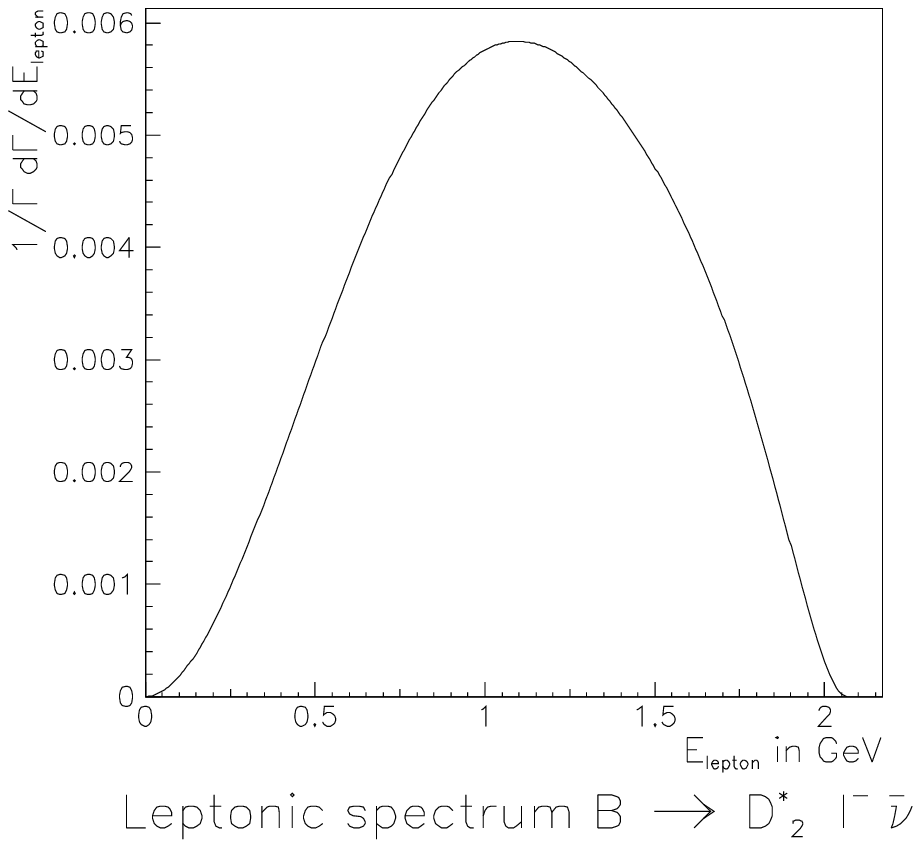}\\
\end{array}$$
\caption{\leftskip 1pc \rightskip 1pc
\baselineskip=10pt plus 2pt minus 0pt {\sl {Leptonic spectra
 $\frac{1}{\Gamma}\frac{d\Gamma}{dE_l}$ of the semileptonic production of
the $j=3/2$ multiplet of $D^{\ast\ast}$ mesons.
}}}
\label{fig:fig5}
\vskip -.3cm\end{figure}
 
In figure \ref{fig:fig3} the leptonic spectrum of the $l=0$
$D$ and $D^\ast$ mesons  are represented.
In figures \ref{fig:fig4} and \ref{fig:fig5} 
the leptonic spectra of the $l=1$ $D^{\ast\ast}$ mesons are plotted for the $j=1/2$ multiplet and for the $j=3/2$
multiplet respectively. We can already
notice two features that will be discussed in the next section:
\begin{itemize}
\item the $j=1/2$ multiplet seems to be outweighed by the $j=3/2$ one
\item the contribution of the $D^\ast_2$ meson is bigger than the 
contribution of the $D_1$ meson
\end{itemize}

\subsection{Comparison with experiment}

\begin{table}[hb]
\begin{center}
\begin{tabular}{|c|c|c|c|c|c|}\hline
Channel & GI$^\ast$ & VD & CCCN$^\ast$ & ISGW & exp. \\ \hline
$B \to D l \nu$ & 2.36 & 2.51 & 2.45 & 1.94  & 1.6 $\pm$ 0.8  \\
$B \to D^\ast l \nu$ & 6.86 & 7.19 & 7.02 & 6.07 & 5.3 $\pm$ 0.8 \\
$B \to D^\ast_2 l \nu$ & 0.70 & 0.47 & 0.65 & 0.77 &  \\
$B \to D_1(\frac 3 2) l \nu$ & 0.45 & 0.29 & 0.42 & 0.49 & \\ 
$B \to D^\ast_1(\frac 1 2) l \nu$ & 0.07 & 0.02 & 0.004 & 0.13 & \\ 
$B \to D_0^{\ast}(\frac 1 2) l \nu$ & 0.06 & 0.02 & 0.004 & 0.11 & \\ 
$\frac {B \to D^\ast_2 l \nu} {B \to D_1(\frac 3 2) l \nu}$&
1.55 & 1.62 & 1.55 & 1.57 & \\
$\rho^2$ & 1.023(3) & 0.98(1) & 0.97(1) & 1.283(1) & \\ 
\hline
\end{tabular}
\end{center}
\caption{\leftskip 1pc \rightskip 1pc
\baselineskip=10pt plus 2pt minus 0pt {\sl {
Semileptonic branching ratios, in percent, taking a $B$ life-time of 
1.62 $10^{-12}$ s.  The last line is the prediction for $\rho^2$. The
column on experiment refers to the rather established and stabilized
results reported in the PDG table. For more recent experimental results,
see the following discussion and table 5. We have starred our preferred models as in table 3.}
}}
\label{tab:larg}
\end{table}

The summary is presented in table \ref{tab:exp}. The agreement is quite
satisfactory for $D^{(\ast)}$.

Concerning the orbitally excited states, some work has to be done to
compare the different experiments between themselves and with our
models.  Experimental numbers are typically provided as branching ratios
$Br(b \to D_J (\to D^{\ast +}\pi^-) l \nu)$ ($D_J$ being any of the two
observed orbitally excited D mesons: $D^{\ast}_2(2460)$ or $D_1(2420)$).
To reduce this to the branching ratio $B\to D_J l \nu$, we use the
nominal production ratio of $\bar b$ jet: $B^+: B^0: B_s: \bar b$-baryon
= 0.4, 0.4, 0.12, 0.08 \cite{pdg2}.

The branching ratios into $D^\ast \pi$ are very different for $D^\ast_2$
and $D_1$. Indeed, the decay $D_1 \to D \pi$ is forbidden by angular
momentum and parity conservation.  We have assumed that these two
channels $D^{(\ast)} \pi$ saturate the $D_J$ decay.  This seems a
plausible assumption, although it has been argued \cite{eichten} that
three-body, $\rho$ and/or $\eta$ decays might be important. This means
that we assume $Br(D_1 \to D^\ast \pi)= 1$, and from isospin
Clebsch-Gordan coefficients, the decay into charged pions $Br(D_1 \to
D^\ast \pi^\pm)= 2/3$. Concerning the $D^\ast_2$ decay, we have two
pieces of information: one from experiment, \cite{cleo} indicates that
$Br(D^{\ast}_2(2460)\to D^+\pi^-)/Br(D^{\ast}_2(2460)\to D^{\ast
+}\pi^-)=2.2\pm0.7\pm0.6$, the other one form theory, the value 2.3 predicted by
HQET \cite{peskin}. Since both agree we take this ratio to be 2.3 implying
$Br(D^\ast_2 \to D\pi)=0.7$ and $Br(D^\ast_2 \to D^\ast\pi)=0.3$, with an
additional C.G. coefficient of 2/3 for final charged pions.

We have treated the above mentioned estimates for the production ratios
$b \to B$ and for the branching ratios $D_J \to D^\ast \pi$ as if they
were exactly known. We do not feel able to discuss in a reliable way the
error attached to them. As a conclusion the experimental errors
mentioned in table \ref{tab:exp} do not incorporate this uncertainty.
They underestimate the error. Of course, every time the experimental
groups have already performed one of the above mentioned steps, we use
their estimate. Then the quoted error is more realistic. For example
CLEO directly provides the $B^-$ branching ratios.

Looking at the table \ref{tab:exp}, it appears that the experimental
numbers are still rather scattered.  The difference in OPAL results
between charged and neutral $D^\ast_2$ is quite surprising.  Our models
agree with the smallest one, $D^{\ast 0}_2$. the other experiments give
only upper
bounds for $D^\ast_2$. None is in clear disagreement with our models. It
is standard to look for the orbitally excited $D$ mesons via their
$D^\ast \pi$ decay. The lower branching ratio of $D^\ast_2 \to D^\ast
\pi$ reduces the number of observed $D^\ast_2$ compared to $D_1$,
allowing only upper bounds, which are not the sign of a small $Br(B\to
D^\ast_2 l \nu)$ but only of a small $Br(D^\ast_2 \to D^\ast \pi)$. This
has to be kept in mind.

The results for $D_1$ are rather  scattered: ALEPH and CLEO
find small values, DELPHI and OPAL larger ones. However the discrepancy
is less than two sigmas. We agree with the small values, ALEPH and
CLEO.

A special comment is due concerning the ratio $Br(B\to D^\ast_2)/Br(B\to
D_1)$. In the heavy mass limit, the two decays depend on one Heavy-quark
universal form factor: $\tau_{3/2}(w)$. Of course, since the kinematic
factors differ between $B\to D^\ast_2$ and $B\to D_1$ (compare eqs.
(\ref{BtoDs2}) and (\ref{BtoD1})), the ratio of the partial widths
slightly depends on the detailed shape of the function $\tau_{3/2}(w)$
which is model dependent. However, this dependence is not very important
as can be seen from the stability of the ratio in table \ref{tab:larg}.  To get some feeling of this dependence we have considered two extreme shapes for $\tau_{3/2}(w)$: a constant and a very steep decrease ($\sigma_{3/2}=3$). The ratio  $Br(B\to D^\ast_2)/Br(B\to
D_1)$ ranges from 1.43 to 1.7.  We may thus
consider a ratio of $\simeq 1.55\pm 0.15$ as a prediction of HQS. Only DELPHI
provides this ratio, which agrees with HQS but within large errors.

\vskip 5mm
\begin{table}[htb]
\begin{center}
\begin{tabular}{|c|c|c|c|c|c|c|}\hline
Channel & Theory & PDG \cite{pdg} & ALEPH \cite{aleph1} \cite{aleph2}&
DELPHI \cite{delphi} \cite{fischer} & OPAL \cite{opal} &CLEO
\cite{cleo2} \\ \hline $B \to D l \nu$ & 1.9 - 2.5 & $1.6 \pm 0.7$ &
$2.35 \pm 0.20 \pm 0.44$ & & &\\ $B \to D^\ast l \nu$ & 6.0 - 7.2 & $5.3
\pm 0.8$ & $5.53\pm0.26\pm0.52$& $5.47 \pm 0.16 \pm 0.67$ & & \\ $B \to
D^{\ast 0}_2 l \nu$ & 0.45 - 0.80 & & $< 0.94$ & $< 6.25$ &
$0.88\pm0.35\pm0.17$ & $< .85$ \\ $B \to D^{\ast +}_2 l \nu$ & 0.45 -
0.80 & & $< 0.54$ & & $2.25\pm0.65\pm0.51$ &
\\
$B \to D_1(\frac 3 2) l \nu$ & 0.30 - 0.50 & & $0.74 \pm 0.16$ & $1.5\pm
0.55$ & $2.0\pm0.6\pm0.5$ & $.49\pm.14\pm.06$\\ $B \to D_1^\ast(\frac 1
2) l \nu$ & 0.0 - 0.07 & & & & &\\ $B \to D_0^{\ast}(\frac 1 2) l \nu$ &
0.0 - 0.06 & & & & &\\ $\frac {B \to D^\ast_2 l \nu} {B \to D_1(\frac 3
2) l \nu}$ & 1.55 - 1.62 & & & $1.55 \pm 1.1 $ & &
\\
\hline
\end{tabular}
\end{center}
\caption{\leftskip 1pc \rightskip 1pc
\baselineskip=10pt plus 2pt minus 0pt {\sl {
The first column gives the range of our theoretical predictions.  We
have assumed a $Br(\bar b\to B^+)=Br(\bar b\to B^0)=0.4$ in order to
compare the experimental numbers referring to $Br(b\to D_J ..)$
with our prediction concerning $Br(B\to D_J l \nu)$. Whenever experiment
gives a branching ratio $\to D_J l X$ we have assumed $X$ to be always a
neutrino. We have furthermore assumed $Br(D_1 \to D^\ast \pi)= 3.33\,
Br(D^\ast_2 \to D^\ast
\pi)$ from the ratio $Br(D^\ast_2 \to D \pi)/Br(D^\ast_2 \to D^\ast
\pi)\simeq 2.3$ and neglecting other $D^\ast_2$ decay channels.
Unable to make a reliable estimate of errors on all these estimates
about $Br(\bar b\to B)$ as well as about the $D_J$ decay branching
ratios, we have chosen to take them with zero error. It results that the
experimental errors on the numbers given in the table are indeed larger
than indicated. The upper bounds are at 95 \% CL, except for the caveat
in the previous sentence.} }}
\label{tab:exp}\vskip 5mm
\end{table}

However, if we take strictly this HQS prediction
 $Br(B\to D^\ast_2 l \nu)/Br(B\to D_1
l \nu)>1.4$, it appears that the central value for $Br(B\to D_1
l \nu)$ in ALEPH and OPAL leads to a prediction for $Br(B\to D^\ast_2 l
\nu)$ which overcomes the upper bound of the same experiment. In view of
the general uncertainty on these numbers, we can only take this as a
faint indication that some of our hypotheses might need to be reconsidered.

Let us first consider the effect of neglecting all decay channels of the
$D^{\ast\ast}$'s except into $D^{(\ast)}\pi$. Let us assume that some
other channels contribute with a partial width to both
$D^\ast_2$ and $D_1$ decays, assumed to be equal to the partial width
into $D^\ast \pi$ times $r$. As a result we should divide in table
\ref{tab:exp} the experimental numbers concerning $D_1$ by $1+r$ while
those concerning $D^\ast_2$ will be divided only by $1+0.3 r$. it
results that this correction would bring the numbers in better
agreement with the HQS prediction $Br(B\to D^\ast_2 l \nu)/Br(B\to D_1
l \nu)>1.4$. At this stage we conclude that it would be prematurate to 
claim that experiment contradicts this HQS ratio. 

Let us still assume that there is some indication of a needed correction
to the strict HQS relation $Br(B\to D^\ast_2 l \nu)/Br(B\to D_1 l
\nu)>1.4$. This leads to consider $O(1/m_c)$ corrections.  A mixing of
the $D_1$ with the $j=1/2$ meson is quite plausible, but it would worsen
our prediction since, due to the very small decay amplitudes into the
$j=1/2$ states, it would lessen our prediction for $Br(B\to D_1 l \nu)$.
We are thus lead to consider direct $O(1/m_c)$ corrections in the decay
amplitudes, which have indeed been found to be dramatically large
in \cite{ligeti}. The authors of the latter paper obtain $Br(B\to D^\ast_2 l \nu)/Br(B\to D_1 l \nu) \simeq 0.8$ for constant $\tau_j(w)$'s.

Unhappily, no direct check of the predicted suppression of the semileptonic
decay into $j=1/2$ orbitally excited $D$'s is presently available. 

\subsection{Comparison with other models}

{\bf Quark models.} Among the calculations of $b \to c$ exclusive semi-leptonic decays which have been performed in the framework of quark models, one must
distinguish between: 1) Calculations in frameworks equivalent to the present B-T formalism \cite{wambach}, \cite{simula}, \cite{simula2}. 2) Calculations  with direct identification of some scaling functions to NR overlaps : \cite{ali}. \cite{SISM}. 3) Calculations
in a family of models with approximate Lorentz boost  \cite{AMT}, \cite{wisconsin}, which are close in spirit to our
old quark model \cite{aleksan}.
4) Calculations which do not deduce the $q^2$ dependence of form factors from wave functions, but from some simple or multiple pole Ansatz ; \cite{heidelberg} (these models have been discussed in \cite{aleksan}), \cite{ISGW2} ; or still some other Ansatz \cite{scora}.

1) {\it Calculations in frameworks equivalent to the present B-T formalism.} In spite of the fact that the authors do not give general formulae,  we have found that the results of ref. \cite{wambach} concide with the ones of the B-T method for the particular case they consider ($m_Q=\infty$ limit, choice of a particular frame, expansion around $w=1$, harmonic oscillator wave functions). Therefore we state that they have
obtained by intuitive arguments essentially the B-T formulation for $m_Q=\infty$. The quite sizeable differences observed with respect to our present quantitative predictions for $\rho^2$ and $\tau_{1/2}(1),\tau_{3/2}(1)$ come from our use of more realistic wave functions: the full numerical solution of relativistic spectroscopic models (using the KE of eq (\ref{kinrel})), instead of the non relativistic ISGW spectroscopic model with harmonic oscillator (h.o.) approximation, yield lower results for $\rho^2$.  

As to the work of \cite{simula}, \cite{simula2}, it is a full calculation of the $B \to D^{(*)} \ell \nu$ form factors with the GI spectroscopic model, in a null-plane approach, which, as we have checked (ref. \cite{raynal}), is the application of the B-T method in the particular $P \to \infty$ frame. As we have suggested \cite{raynal}, because of the demonstrated covariance of the $m_Q=\infty$ limit, the results should then coincide in this limit with our own, provided the same spectroscopic model is used. Indeed our $\xi(w)$ agrees within the computational uncertainty with the prediction by these authors when using the same GI model.

2) {\it Calculations  with direct identification of some scaling functions to NR overlaps}. In \cite{ali}, \cite{SISM}, a simple intuitive recipe has been used, consisting in the direct identification of the scaling functions of Isgur and Wise with usual non-relativistic form factors. It results in quite
different expressions from what we advocate througout this paper: deducing the scaling functions from their relation to the current matrix-elements, the latter being calculated in a relativistic approach, as is  proposed by the B-T approach and the one below (point 3)). 

3) {\it Calculations
in a family of models with approximate Lorentz boosts.} In \cite{AMT}, \cite{wisconsin}, the approach is essentially the one of our old model developped in the 70's (in the h.o. case, one ends with exactly the same $\xi$ that we had found; for a recent reference, see \cite{aleksan}). It consists in calculating the current matrix elements between wave functions for states in motion obtained through a Lorentz transformation of rest-frame wave functions, including the Lorentz transformation of coordinates and the standard transformation of spinors in Dirac representation. The matrix elements are calculated in the equal-velocity frame where great simplifications occur. To get the equal time wave function from the one at rest, one has to assume that the dependence on the relative time is weak in this latter frame (this corresponds to the assumption of factorisation
of the wave function in light and heavy degrees of freedom in \cite{AMT}, \cite{wisconsin}. One then ends with what we have called the ``na\" \i ve Lorentz contraction''
prescription for spatial wave functions. This prescription can be seen to be too inaccurate to calculate form factors. In the B-T formalism, there appears a definite and more consistent Lorentz transformation of momenta. The 
 ``na\" \i ve Lorentz contraction'' prescription consists in peforming in the result of the BT boosts a linear approximation in the dependence on the light quark spatial momenta $\vec p_2$ in the hadron rest frame .
The resulting difference can be seen in the behavior of the function $\xi(w)$. With the simpler Lorentz contraction prescription, $\xi(w)=2 I(w)/(1+w)$ where
the overlap factor $I(w)\to$ constant when $w\to\infty$\footnote{In the harmonic oscillator case $\xi(w) = \xi_{NR}(w)$ (see eq. (\ref{NR}))}. Hence  
$\xi(w) \sim (2/(w+1))$ for large $w$. 
In the B-T formalism, as we have argued after eq. (\ref{delta}), $\xi(w) \simeq (2/(w+1))^{2 \rho^2}$, with a power of dynamical origin.

Another point is deserving mention as regards models \cite{AMT}, \cite{wisconsin}. 
Within a static approximation where the light quark wave functions at rest would have no small Dirac components ("non relativistic" case of \cite{wisconsin}), and assuming, as is found to a good accuracy, that the spin-orbit effect is small, one gets $\tau_{1/2}(1) \simeq \tau_{3/2}(1)$. On the other hand, if one uses wave functions with small components, like for instance solutions of a Dirac equation ("DESC" of \cite{wisconsin}), one has $\tau_{1/2}(1) \not = \tau_{3/2}(1)$, and indeed, one finds, similarly to our B-T result, $\Gamma(1/2) << \Gamma(3/2)$. In fact the B-T formalism corresponds to having small components given by the free-quark Dirac spinor structure.

Apart from particular features, one must emphasize the {\bf theoretical advantage of the B-T approach} that comes from the {\bf demonstration of  exact general properties} such as {\bf unitarity} of the transformation from rest frame  to states in motion, or, in the $m_Q=\infty$ limit, {\bf Isgur-Wise scaling, full covariance and duality}. In the other approaches, either they do not hold, or they are enforced by formal procedures, for example formal covariance can be obtained by identifying invariant form factors with form factors in one particular frame like the equal-velocity frame, or they are only approximate.

Finally, one must still mention the following
important point: the above discussion has been maintained within the $m_Q=\infty$ limit. For our B-T approach, as well as for the others, it is however quite possible to treat the finite mass case, and it is indeed one interesting feature of quark models that they yield definite $1/m_Q$ effects. We have only refrained to do so because our approach loses several of its nice properties, in particular covariance.
Detailed predictions for finite $m_Q$ are given in the ISGW2 model \cite{ISGW2}. This model falls in the category 4), in that {\it the form factors do not really derive from a full quark model calculation}. 
 One interesting finding of the ISGW2 calculations \cite{ISGW2} is that they predict $\Gamma(2^+) < \Gamma(1_{3/2}^+)$, in contradiction with HQS. This may be due to the finite mass treatment and would be in agreement with the conclusions of \cite{ligeti}. It may also be mentioned  that for the ISGW1 model (\cite{scora}), $\tau_{3/2}(1)$ is sensibly smaller than ours (0.3 against 0.55).

{\bf QCD fundamental methods.} By fundamental methods, wee mean lattice QCD and QCD sum rules. As to the comparison of their results with ours as regards the elastic Isgur-Wise function $\xi(w)$, the discussion has been made in  \cite{rho2}, and the convergence is encouraging, even as concerns the sensitive $\rho^2$ parameter. For orbitally excited states, only QCD sum rules give predictions (\cite{colangeloSR}), and they lie
well below the ones of our quark models for ununderstood reasons (with $m_Q=\infty$, $Br(B\to D_1{(3/2)}^+ l\nu) =0.1\%$ versus $0.3-0.5\%$ in table \ref{tab:exp}).

\section{Conclusion}

We have applied the Bakamjian-Thomas (BT) method in the infinite mass limit to 
meson wave functions fitted to the experimental meson spectrum by several groups: GI, VD and CCCN, \cite{godfrey}-\cite{cea}, using a relativistic kinetic term and ISGW \cite{scora} a non relativistic one. Whichever set of wave function is used, 
the BT method insures that the models are covariant and satisfy Heavy Quark Symmetry
\cite{raynal}-\cite{dstar}. 
 
We have computed the invariant form factors $\xi(w)$ and $\tau_j(w)$.
The Bjorken sum rule, which has been proven to be valid in BT models \cite{duality}, \cite{dstar}, has been checked practically in this paper. The 
derivative Bjoren sum rule for $w= 1$ is almost saturated by the ground state $l=1$ and the first $l=1$ radial excitation (the latter contributing for $\simeq 15\%$), except for the VD model, in which $\sim 20\%$ is contributed by the $n>1, l=1$ excitations. The full Bjorken sum rule has also been checked
for $1\le w \le 1.5$. The missing part, corresponding to the $l >1$ states, increases with $w$ up to 15\% for $w=1.5$ (25\% for the ISGW model).

 The slope $\rho^2$ is $1.\pm 0.05$ when relativistic kinetic energy is used, $\simeq 1.3$ when the non-relativistic kinetic energy is used \cite{rho2}. The $\xi(w)$ function is very well fitted by the function $(2/(w+1))^{2\rho^2}$ which means a dipole like behaviour, eq. (\ref{dipole}). The latter result may be traced back to the Coulomb-like shape of the ground state wave function when using a relativistic kinetic energy \cite{rho2}.
The functions $\tau_j$ are dominated by $\tau_{3/2}$, with $\tau_{3/2}(1)\simeq 0.4 - 0.6$. The latter decrease like $(2/(w+1))^3$.

We have computed, for the decays $B\to D l\nu , D^\ast l\nu$ and the four $B\to D^{\ast\ast}l \nu$, the differential decay widths as a function of $w$, figure \ref{fig:fig6},  and also as a function of the charged lepton energy. Figures \ref{fig:fig3}-\ref{fig:fig5} show the latter. The decay widths to $j=3/2$ are about one order of magnitude smaller than the ones into $l=0$ states, and the ones into $j=1/2$ are still one order of magnitude below. 

We have also computed the semileptonic branching ratios. The ones into $D$ and $D^\ast$ are in good agreement with experiment. For the $D^{\ast\ast}$ experimental data are still rather scattered. Our models predict tiny branching ratios into $j=1/2$ states,  but this cannot be checked yet. For sure these decays have not been seen yet, but their absence may be explained by their broad widths, without invoking a suppressed branching ratio. Our predictions for $D^\ast_2$ are below the experimental upper bounds, although
close to them. Our predictions for the $D_1(j=3/2)$ do not disagree with experiment taking into account the wide spreading and large uncertainty of experimental results. 

Our models being all within the strict infinite mass limit, we predict the relation 
$Br(B\to D^\ast_2)/Br(B\to D_1(j=3/2) =1.58 \pm 0.05$ which is  
mainly a consequence of HQS. Indeed, going beyond the models studied in this paper, if we consider arbitrary but
``reasonable'' shapes of $\tau_{3/2}(w)$ we find that HQS predicts  $Br(B\to D^\ast_2)/Br(B\to D_1(j=3/2) =1.55 \pm 0.15$. When comparing the experimental figures for these decays, there might be some indication that this ratio is indeed lower. This could be an indication of a large $1/m_c$ correction which enhances the $D_1$.

\section*{Acknowledgements}

It is a pleasure to thank Pierre-Alexandre Fischer, Dominique Pallin, 
Patrick Roudeau and Achille Stocchi
for precious informations and interesting discussions.

\end{document}